\documentclass
	[a4paper,twocolumn,twoside,10pt,showkeys]
	{revtex4}

\usepackage{amsmath,amssymb}	
\usepackage{bm}					
\usepackage{color}					
\usepackage{graphicx}				
\usepackage{dcolumn}				

\newcommand{\figCapSkip}{\vspace{-4ex}}	
\makeatletter
\def\frontmatter@abstractwidth{0.9\textwidth}	
\makeatother

\begin{document}



\newcommand{\By}{$\times$}
\newcommand{\SqrtBy}[2]{$\sqrt{#1}$\kern0.2ex$\times$\kern-0.2ex$\sqrt{#2}$}
\newcommand{\Degree}{$^\circ$}
\newcommand{\DegreeC}{$^\circ$C}
\newcommand{\Ohmcm}{$\Omega\cdot$cm}

\title{
Classification of Light-Induced Desorption of Alkali Atoms \\
in Glass Cells Used in Atomic Physics Experiments

}


\author{Atsushi Hatakeyama}
\email[Corresponding author: ]{hatakeya@phys.c.u-tokyo.ac.jp}
\affiliation{%
Department of Basic Science, Graduate School of Arts and Sciences, University of Tokyo, Komaba, Tokyo 153-8902, Japan
}

\author{Markus Wilde}
\affiliation{%
Institute of Industrial Science, University of Tokyo, Komaba, Tokyo 153-8505, Japan}

\author{Katsuyuki Fukutani}
\affiliation{%
Institute of Industrial Science, University of Tokyo, Komaba, Tokyo 153-8505, Japan}

\begin{abstract}

\vspace*{1mm}

We attempt to provide physical interpretations of
light-induced desorption phenomena that have 
recently been observed
for alkali atoms on glass surfaces of alkali vapor cells used in atomic physics experiments. 
We find that the observed desorption phenomena are closely related to
recent studies in surface science, and can
probably be understood in the context of these results. 
If classified in terms of the photon-energy dependence, the coverage and the bonding state of the alkali 
adsorbates, the phenomena fall into two categories:
It appears very likely that the
neutralization of isolated ionic adsorbates 
by photo-excited electron transfer
from the substrate
is the origin of the 
desorption induced by ultraviolet light
in ultrahigh vacuum cells.
The desorption observed in low temperature cells, on the other hand, which is
resonantly dependent on photon energy in the visible light
range,
is quite similar to light-induced desorption 
stimulated by localized electronic
excitation on metallic aggregates.
More detailed studies of light-induced desorption events from surfaces well characterized
with respect to alkali coverage-dependent ionicity and aggregate morphology appear highly desirable for the development of more efficient alkali atom sources suitable to improve a variety of atomic physics experiments.
\end{abstract}


\maketitle
\newpage


\section{Introduction}

Alkali-atom desorption stimulated by photon irradiation is
an interesting subject in atomic physics experiments, and often
called ``light-induced atom desorption'' (LIAD).
Alkali atoms are widely used in atomic physics experiments because of their
simple structure and the availability of handy laser sources such as
laser diodes for exciting these atoms.
Because alkali metals, in particular
heavier alkalis like K, Rb and Cs, have
relatively high vapor pressures, evacuated and sealed glass cells containing bulk alkali metal
are convenient samples that provide sufficient atomic densities 
in the gas phase at room temperature and above
for laser spectroscopic experiments.
Therefore the sealed vapor cells have long been basic tools in atomic physics laboratories, and are currently being utilized as well
in many advancing experiments.
Only a few of them are listed here, with the employed alkali atoms mentioned
in parentheses: Ultrasensitive magnetometer (K)~\cite{Kom03};
frozen light pulse (Rb)~\cite{Baj03};
chip-scale atomic clock (Cs)~\cite{Kna04}.

In the conventional vapor cell, the atomic density
in the gas phase can be increased by
heating the cell, which raises the vapor pressure.
However this method is not desirable or not applicable at all
for some recent experiments, in which LIAD from the glass surfaces of
cells is exploited to increase the atomic densities
quickly on demand.
One of such cells is an ultrahigh vacuum (UHV) cell used in laser
cooling and trapping experiments. These UHV cells are continuously evacuated
with vacuum pumps to prolong the trap lifetime by reducing
collisions with background gases. 
The typical cell pressure
is 10$^{-8}$~Pa, which is much lower than alkali vapor pressures at
room temperature (10$^{-5}$~Pa (10$^9$~atoms/cm$^3$) 
for Rb). 
LIAD of K and Rb from borosilicate
glass (Pyrex and Vycor) surfaces has been observed in
these cells by irradiation of ultraviolet (UV)
photons~\cite{Han01,Du04,Aub05,Nak05,Kle05}. This phenomenon is useful
especially for experiments generating Bose-Einstein condensations
(BECs), which require seemingly contradicting conditions of 
both large numbers of trapped atoms (i.e. a high density of sample atoms)
and long trapping times (i.e. low untrapped gas pressures).
The recent achievement of a compact BEC system using a microelectronic
chip~\cite{Han01} owes its success partly to this LIAD.

The other type of cells in which LIAD is observed are low
temperature (LT) cells, sealed
alkali (Rb, K) vapor
cells (containing He buffer gas) 
cooled down to liquid helium temperatures 
($\sim 2$~K)~\cite{Hat00, Hat02, Hat04}.
The observed LIAD, which exhibits a resonant dependence on the photon energy
in the visible light range, is used to disperse alkali
atoms effectively into a cryogenic helium gas. It was then demonstrated
that the cryogenic helium gas environment preserves quite long
the spin polarization of the ground state of alkali atoms. 
This demonstration has attracted much attention~\cite{Bud02} 
because the slow spin relaxation of the
alkali ground state is always a key in many experiments
that exploit the ground state polarization and coherence,
including the above mentioned that so far use conventional vapor cells.

In spite of the important contributions of these LIAD phenomena
to atomic physics, however, 
the understanding of their mechanisms is quite limited, and
the LIAD conditions have been optimized only
empirically. Recently, in surface science, two studies have
been reported which seem quite helpful to understand desorption
mechanisms in the vapor cells.
One study was motivated by the search for the origin of
atomic Na and K in the planet Mercury and the Moon~\cite{Mad02}.
These celestial bodies have tenuous atmospheres containing atomic Na and
K, which must be continuously supplied for the steady content of the alkalis.
Photon-stimulated desorption was proposed as the origin
of the alkali atoms,
and laboratory experiments~\cite{Yak99,Yak00,Yak03} 
demonstrated effective desorption induced by
UV photons from model
mineral surfaces (amorphous SiO$_2$ surfaces). 
It is supposed that the desorption is caused by charge transfer excitation
from SiO$_2$ to the unoccupied orbital of the almost completely cationic
alkali atoms bound to certain surface defects~\cite{Dom04}.
The other study investigated desorption from alkali (Na and K)
nanoparticles (an effective mean size of 10~nm) formed on a quartz substrate held at liquid nitrogen temperature~\cite{Mar03}.
This desorption was found to be resonantly dependent on the photon energy in the visible
light region. It is claimed that the localized electronic excitation of
certain binding sites is responsible
for the resonant desorption, and that surface plasmon resonances of the nanoparticles
may enhance the desorption rate.
This mechanism may be used to manipulate metal surfaces on the atomic
scale.

The purpose of this paper is to reexamine the LIAD phenomena observed
in atomic physics experiments on the basis of the accumulation of
desorption models developed in the field of surface science.
We find that the observed desorption phenomena seem to be closely related to
the two surface science studies mentioned above, and can
probably be understood in the following context: it
appears very likely that desorption in the UHV cells originates from
the neutralization of ionic isolated adsorbates 
by photo-excited electron transfer
from the substrate; while desorption in the LT cells seems consistent
with LIAD stimulated by local
resonant excitation on metallic aggregates.
However
we have also noticed that the comprehensive study of desorption
for various forms of adsorbates (from isolated atoms to aggregates)
on well characterized surfaces
under illumination of photons in a wide energy range (from infrared to UV)
is still missing.
Such a study would provide new insight for
the development of a new type of efficient atomic source that can be
useful in a variety of atomic
physics experiments.

It is noted that LIAD in vapor cells was first observed
for silane coatings
(in particular, polydimethylsiloxane) on cell walls in
the early 90s~\cite{Goz93}, and since then there have been many investigations~\cite{Meu94,Xu96,Atu99,Atu03}
including desorption from a paraffin coating~\cite{Ale02}.
In this paper we do not discuss LIAD from polymer coatings, because
its mechanism is probably different from the desorption from bare glass
surfaces. A recent study on this subject is reported in Ref.~\cite{Bre04}. 

\section{Review of LIAD in alkali vapor cells}

\subsection{Ultrahigh vacuum cells}

Several groups have reported LIAD in UHV
cells~~\cite{Du04,Aub05,Nak05,Kle05}. Although each
group performed the experiment in a slightly different way,
we summarize here the general properties of LIAD in UHV cells.
The UHV cell is kept evacuated by vacuum pumps, usually ion pumps, 
at 10$^{-7}$ - 10$^{-8}$~Pa.
It is usually baked at about 150{\DegreeC} at
the cell preparation stage. 
Alkali atoms
are supplied into the cell by heating an alkali 
metal dispenser~\cite{For98} or reservoir when necessary.
A typical procedure of filling the cell with Rb atoms is as
follows~\cite{Hor05}: A dispenser is heated for
15 minutes, while the pressure indicated by the ion pump current
remains below $10^{-7}$~Pa. After termination of the heating, the cell
is left overnight, during which the pressure of the cell returns
to the stationary value (10$^{-8}$~Pa) due to the pumping of 
the alkali atoms and other gases by the cell surface and the ion pump.
Laser trapping experiments are performed on the following day
using LIAD from the alkali-impregnated (but transparent) cell surfaces to increase the number of Rb atoms in the gas phase.
It should be noted that the desorption efficiency is not dramatically
reduced even if the cell is left for one month after heating the dispenser.
A similar but longer-cycle heat-and-wait procedure is 
reported in Ref.~\cite{Kle05}.

The irradiation of the cell walls
with relatively high energy photons
from, e.g., halogen bulbs or blue LEDs (photon energy: $\sim$3~eV)
rapidly induces desorption, which is monitored by the
increase of the number of laser-trapped atoms.
Although the absolute number of trapped atoms depends on
the configuration of the laser trapping system, this number
is raised by two orders of magnitude to $10^9$ atoms
with the LIAD light (blue LEDs of
140~mW, 3.1~eV) on~\cite{Aub05}, indicating that
the atomic density in the gas phase increases by a similar order.
The number of trapped atoms increases with increasing photon energy
in the range of 1.6~eV to 4.9~eV~\cite{Kle05}.
After a sufficient number of atoms are trapped, the cell pressure is
reduced quickly by turning off the LIAD light. 
Then the trapped atoms are processed further for, e.g., generating BEC.

Rb is the most popular sample in BEC experiments,
and its desorption has been reported by many groups 
from Pyrex glass~\cite{Du04,Aub05,Nak05}, one of
the most commonly used material for glass cells (typical composition: SiO$_2$ 81\%;
B$_2$O$_3$ 13\%; Na$_2$O 4\%; 
Al$_2$O$_3$ 2\%), and Vycor glass 
(SiO$_2$ 96\%;
B$_2$O$_3$ 3\%; Al$_2$O$_3$ 1\%)~\cite{Kle05}.
K desorption is also reported for Pyrex~\cite{Aub05} and Vycor~\cite{Kle05}.
It is reported that quartz is much less suitable for LIAD than Pyrex~\cite{Du04}.

It is noted that in UHV cells desorption from metal surfaces
was also observed for Rb on stainless steel~\cite{Kle05,And01}
and Cs on aluminum~\cite{And01}.

\subsection{Low temperature cells}

LIAD has been observed by one of
the present authors in LT cells, sealed Rb vapor cells cooled
to liquid helium temperatures.
These glass (Pyrex) cells
are filled at room temperature in a
manner similar to conventional sealed glass cells (see, 
for example, Ref.~\cite{New93}). 
The cells are prepared under high vacuum conditions ($\sim10^{-4}$~Pa), 
baked at a few hundreds degrees of Celsius for a day.
Rb metal is transported 
into the cell from a Rb reservoir 
through a glass stem connected to the main
body of the cell by heating the metal
with a hand torch.
The empirically best way to maximize the desorption efficiency
is to cover the cell surface once with a visible Rb film and then remove
it by heating the cell just until the film becomes invisible
with the naked eye. Desorption is not observed from visible Rb films
on the surfaces. The cell is filled with high-pressure
helium gas (typically 3.6~atm at room temperature),
which works as a buffer gas to prevent frequent collisions of Rb atoms
with the surface, and then
sealed by melting off the glass stem.
 No visible Rb metal is left in the main body of the cell, while
 a small amount of solid Rb exists in the remaining part of the stem.
The vapor pressures of Rb in the cells vary from cell to cell
and usually amount to several ten percent of the saturated vapor
pressure. This is probably because bulk Rb metal is almost confined to
the stem, and the surface of the main body itself is not covered by metallic Rb~\cite{Zen85}.
This kind of undersaturation is often observed in sealed vapor cells
that do not contain a large reservoir of uncontaminated bulk metal.
Cells prepared in a similar fashion in a different laboratory also
show LIAD~\cite{You99}. Thus we believe the observed
LIAD in the LT cells is generally reproducible by
preparing cells in this way.

\begin{figure}[tbh]
\begin{center}
\includegraphics[width=5.5cm]{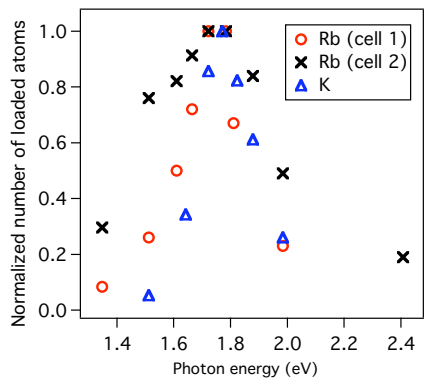}
\end{center}\figCapSkip
\caption{\label{fig1}
Photon-energy dependence of the number of atoms loaded by LIAD at 1.85~K
for two Rb cells and one K cell. The data are normalized to each peak value.}
\end{figure}

The cell is gradually cooled from room temperature.
The number of gaseous Rb atoms, which is monitored by observing
fluorescence of atoms
excited by a probe laser, decreases with decreasing temperature because
the saturated pressure of Rb decreases and the gaseous
Rb atoms adsorb
on the cell surface. At about 250~K, the atomic density
of gas-phase alkali atoms falls below the limit of facile detectability
($\alt 10^7$~atoms/cm$^{3}$).
The loading of Rb atoms into the dense helium gas (10$^{20}$
~atoms/cm$^3$) in the cell by LIAD
then becomes dramatically
effective below the temperature at which the helium gas
begins to liquify and at the same time
becomes superfluid (typically 1.9~K).
This is supposed to be due not
to an increase of the desorption efficiency, but
to the effective transport of desorbed
atoms from the surface into the dense helium gas by the gas
flow induced by the evaporation of the superfluid film (see
Ref.~\cite{Hat02} for details).
The typical density of loaded atoms is 10$^8$~atoms/cm$^3$ for
a cw desorption laser (50~mW/cm$^2$, 1.8~eV).
In Fig.~\ref{fig1}, the photon energy dependences of
Rb loading
in different cells at 1.85~K are shown.
The energy dependences differ slightly
from cell to cell, particularly in their widths, but
generally they peak at about 1.75~eV. This resonant behavior
is quite different from the energy dependence observed in the
UHV cells, where the desorption efficiency decreases
with decreasing photon energy down to 1.6~eV.
K desorption is also observed in K cells. 
It seems to occur at slightly higher energy than Rb
desorption (see Fig.~\ref{fig1}). 

The loading efficiency decreases upon continued desorption
light irradiation. It can be recovered by heating the cell to about
200~K or above for a while, and subsequently cooling it down again.

The loading of atoms by desorption is barely observable at temperatures 
above the superfluid transition temperature in the cell
up to about 250~K, 
at which residual Rb vapor hinders the detection of desorption. 
There seems to be no
strong temperature dependence of the number of loaded atoms.
 
\section{Discussion}

The desorption observed in the UHV cells is quite similar,
in terms of the photon energy dependence,
to the desorption studied using electron impact as well
as photon irradiation in the context of the search for the origin of
atomic alkalis in tenuous planetary atmospheres~\cite{Yak99,Yak00,Yak03}.
In these studies alkali desorption from model
mineral surfaces (amorphous SiO$_2$)
exhibits a threshold at $\sim$3~eV
for K and $\sim$4~eV for Na,
a local maximum at 9-11~eV for both K and Na, and a general trend of increasing desorption rate with increasing photon or electron energy.
According to a theoretical modeling~\cite{Dom04}, the threshold can be
explained by desorption of alkali adsorbates at non-bridging oxygen (NBO) defects due to
the local charge transfer from the O(2p) nonbonding orbitals to
cationic adsorbates, while
the resonant-like peak at 9-11~eV can be related to a direct or indirect excitation
from the O(2p) valence band to the cations.
Considering the photon energies used for the desorption in the UHV
cells, this desorption would be attributable
to the former process.
The cross section of the Na desorption at 5~eV is reported to be about
$3 \times 10^{-20}$~cm$^2$~\cite{Yak99}. If we take this value for
desorption observed in the UHV cells and assume the surface density
of the desorbable atoms to be 10$^{11}$~cm$^{-2}$, the desorption rate
is estimated to be about $10^6$~cm$^{-2}$s$^{-1}$ for desorption
light of 1~mW/cm$^2$.
This estimation appears to be quite reasonable~\cite{Kle05}.

For the desorption observed in the LT cells,
the idea that it originates from alkali clusters is not inconsistent
with regard to the resonance-like photon energy dependence in the visible
light range~\cite{Mar03}, which strongly resembles the desorption from alkali clusters.
This resonant behavior has long been supposed to
be related to the surface plasmon resonances of alkali clusters~\cite{Hoh93}.
However,
a recent study points out that this kind of resonant desorption may also result from
the localized resonant excitation of certain bonding sites of alkali atoms on the alkali cluster surface and the desorption described in Ref.~\cite{Mar03} was attributed to such a mechanism~\cite{Jac02}.
Note that this desorption scenario has its origin in the excitation 
to the 
lowest $p_{3/2}$ states of
atomic alkali in vacuum (1.62~eV for K and 1.59~eV for Rb). 
The photon energy dependence (see Fig.~\ref{fig1})
observed in the LT 
cells for K is strikingly similar to that observed in Ref.~\cite{Mar03} for
K clusters. Thus the mechanism of resonant desorption in the LT
cells can either be of this site-selective resonant type, or
plasmon-assisted desorption as discussed in Ref.~\cite{Hat02}.

In order to further examine the validity of
the above two explanations,
it is essential to characterize the adsorbate morphology of
the vapor cells
as already pointed out by one of the present authors~\cite{Wil99}.
Although the two classes of desorption phenomena mentioned above are observed for
K and Rb atoms adsorbed on bare glass surfaces~\cite{comment1}, their characteristic photon
energy dependencies are distinctly different, which points towards differences in the 
desorption mechanism operative in each case. 
One important factor affecting the desorption process is probably the difference in the 
surface condition of the alkali adsorbates as present in the two types of cells. 
However, the surfaces of the vapor cells used in the above experiments
are not carefully prepared and characterized with respect to the alkali coverage and
its aggregate condition, as possible e.g. by standard surface characterization techniques such as X-ray photoelectron spectroscopy (XPS) and work-function measurements.
Therefore we here try to estimate the surface conditions in the
UHV cells and the LT cells on the basis of a comparison to surface studies performed under similar experimental conditions.

There are many reports of alkali atom adsorption on
oxide surfaces including SiO$_2$, the main component of common glass.
From these studies, it is known that
defect sites of the SiO$_2$ surface such as NBO
centers strongly adsorb alkali
atoms, while the bonding to regular sites is weak~\cite{Bra97,Lop99}.
The binding energies to such defect sites 
are calculated to range from 1 to 3~eV for Cs~\cite{Lop99}.
Similar binding energies are suggested experimentally 
for Na~\cite{Yak00_2}.
It is worth mentioning that
adsorption with such a high adsorption energy
is often considered as ``reaction" or ``loss"
of alkali atoms on the surface in atomic physics experiments
using conventional vapor cells~\cite{Ste94}
because once adsorbed the atoms do not
return to the vapor phase by moderate elevation of temperature~\cite{Yak00_2}. Note that penetration of alkali atoms
into glass may occur~\cite{Yak00_2}.

Alkali atoms adsorbed to defects
are present as almost completely ionized
cations by transferring their valence electron to the 
surface~\cite{Dom04,Bra97,Lop99,Yak00_2}.
The adsorbed ionic atoms are repulsive to each other.
The surface density of this kind of adsorbate depends on that of the
defect sites, which is affected by
the processing history of the glass surface
as well as by the type of the glass. Although most defect sites would be occupied
by various species such as carbonates and hydroxides~\cite{Him97}
before cell preparation due to, for example, exposure to air,
it is reasonable to assume that some of these defect sites would become
unoccupied and reactive to alkali atoms during the evacuation and baking
processes. 

The surface continuously exposed to alkali atoms
as in vapor cells will finally become ``non-reactive" (``cured")~\cite{Ste94}, with all high binding energy sites saturated
by adsorbed alkali atoms. On such a cured surface, the adsorption energy 
for additional adsorbates is lowered, measured using
conventional vapor cells to amount to 0.53~eV for Cs on Pyrex~\cite{Ste94} and 0.66~eV on fused silica~\cite{Bou99}.
These adsorption energies are close to the heat of vaporization
of bulk Cs (0.67~eV), and therefore the adsorbed alkali may
 become
metallic under these conditions. 
This transition of the alkali morphology from ionic to metallic 
with increasing coverage is reported in many surface science
 studies~\cite{Bra97, Yak00_2}.
 The surfaces of the UHV cells are probably quite similar to such cured vapor cell surfaces. It should be remarked, however, 
that 
we do not consider that metallic aggregates are present
 on the surface 
because the partial pressure of alkali is much lower than the
saturation vapor pressure in the UHV cells at room temperature
 and thus metallic particles are unstable~\cite{Bra97}.

The density of atoms weakly adsorbed to the cured surface
is estimated using the following relation~\cite{Boe68} about the residence time on the surface
$\tau$:
\begin{equation}
\tau = \tau_0 \exp (E_a/kT),
\end{equation}
where $\tau_0$ is on the order of
$10^{-12}$ -- $10^{-14}$ s, $E_a$ the adsorption
energy, $k$ Boltzman's constant, and $T$ the surface temperature.
By balancing the numbers of atoms impinging on and leaving
the surface, one can calculate the surface density.
Even for a relatively large adsorption energy of 0.7~eV and a long $\tau_0$
of 10$^{-12}$~s, the surface density of weakly adsorbed atoms is still
estimated to be low (compared to the monolayer coverage density of
$10^{14}$~cm$^{-2}$), i.e. 
10$^{10}$~cm$^{-2}$ in a UHV cell with an alkali vapor
pressure of 10$^{-8}$~Pa.
This estimation also supports the picture that
the adsorbed atoms
on the UHV cell surfaces are mainly isolated. Most of them are strongly
bound to defects, while others are weakly trapped and 
eventually return into the gas phase.
These conditions are similar to the
ionic adsorbate conditions of low coverage surfaces in the studies of
Na and K desorption from amorphous SiO$_2$.

The situation of the LT cells is different. Their surfaces
are usually once covered with alkali films at the filling stage
(although the films are removed afterwards by heating
just until they become invisible with the naked eye), and
are kept exposed to alkali vapors at pressures nearly equal to the
saturated vapor pressures.
These are favored conditions for the presence of alkali aggregates
on the surface. The cooling of the cells is also favorable for
aggregate formation, because the surface density of adsorbates
increases. It is known that
at low temperatures additional exposure to vapors can lead
to formation of metallic aggregates~\cite{Gra97,Yak00_2,Wil99}. 
Therefore it is
rather likely that in the LT cells there exist some alkali aggregates or clusters on the surface, similarly to the study of desorption from
alkali nanoparticles formed on the quartz surface.
However, it should be pointed out that a solid conclusion must wait
for the experimental characterization of the cell surfaces especially
because we do not precisely know the amount of deposited alkali metal.

As seen above, the estimation of the surface conditions also 
supports close relations between the following desorption
 phenomena: 
desorption observed
 in the UHV cells and desorption of ionic adsorbates 
from amorphous SiO$_2$ on one hand, and desorption observed 
in the LT cells and desorption from 
alkali nanoparticles formed on quartz on the other. 
Thus, it appears reasonable to conclude that the desorption observed
in the UHV cells originates from
 the neutralization of ionic alkali atoms adsorbed on defects 
by photo-excited electron transfer
from the substrate, whereas the desorption observed in the LT cell may 
occur from alkali metal aggregates. 

\section{Future perspectives and conclusion}
Light-induced
desorption of alkali atoms deposited on glass surfaces
may occur via various mechanisms in a wide photon energy range
from different forms of adsorbates, 
and the desorption phenomena observed
in the UHV cells and the LT cells are only two prominent examples.
If their mechanisms could be clarified, there should exist many possibilities to increase
the desorption efficiency
by optimizing the experimental conditions.
For this purpose, it is very desirable to study the desorption process
of ionic (isolated) and metallic (aggregated) adsorbates
under well-characterized adsorption conditions 
using various photon energies from infrared to ultraviolet.
The surface characterization with respect to the initial state of the desorption event in terms of the alkali coverage and bonding conditions is quite critical to derive a solid
conclusion. Not only the photon energy dependence but also
the analysis of the kinetic energy distribution of the desorbing atoms
should be very helpful to elucidate the desorption mechanisms. 

Experiments that nearly fulfill these requirements have been
reported for K on Cr$_2$O$_3$ surfaces~\cite{Wil99} as well as
for Na and K on SiO$_2$ surfaces~\cite{Yak99,Yak00,Yak03}. 
Desorption from both ionic
and metallic adsobates is studied, although only high energy photons
($\agt$ 3~eV) were used. The result shows that the desorption cross section is
higher for ionic adsobates than for aggregates at this 
photon energy. However, because of many possible desorption sites
in metallic aggregates at larger alkali coverages, the total desorption rate is not necessarily smaller than in the case of ionic atoms at much lower coverage. It is further
interesting to point out that for large-scale metallic adsorbates
 the desorption rate may even increase under prolonged light irradiation, which indicates that the desorption cross section may be strongly sensitive to the metal particle morphology and/or the total alkali coverage. 

From these as well as the other results described in this
paper, the following can be anticipated.
For desorption of isolated ionic atoms, which are supposed to be present as majority species on low-coverage surfaces in the UHV cells, it seems promising to use as substrates glass surfaces that have many active
defects sites. Mechanically activated surfaces like the surfaces
of cleaved or fractured glasses under UHV conditions are hopeful
candidates~\cite{Rad95}. Higher photon energy appears preferable, regarding the general trend of increasing desorption efficiency towards large excitation energies.
Aggregates prepared on a high-coverage surface cooled in UHV chambers
could be an effective LIAD source. It would be helpful to see which
photon energy is the most efficient for aggregates of a given size, or whether in turn the aggregate size could be controlled to yield particularly efficient desorption at a certain excitation energy.

In the case of the LT cells, where the surface condition is
less well understood than that of the UHV cells, 
it is important to identify
first the morphology of adsorbates from which desorption occurs.
This would be the first step to understand the desorption mechanism
and develop a better atom loading method.
It is noted that desorption using UV light has not
been attempted yet and is therefore worth trying.

In conclusion, on the basis of models that have been developed
in the surface science field,
we reexamined LIAD phenomena recently observed
in atomic physics experiments.
Empirically, the reviewed results may be classified into two categories of LIAD phenomena, i.e. as desorption stimulated by UV light in 
UHV cells on one hand, and resonant desorption in the visible light range
in LT cells, on the other.
It is very likely that desorption in the 
UHV cells is induced by the
neutralization of the ionic isolated adsorbates 
by photo-excited electron transfer
from the substrate. The mechanism of this type of desorption
has been studied in a series of studies motivated
by the search for the origin of atomic alkalis in planetary
atmospheres. The desorption in the LT cells
appears to be closely related to LIAD 
from metallic aggregates by localized resonant excitation. 
These conclusions were supported by estimation of
the surface conditions of these two types of
cells: Isolated and ionic adsorption of alkali
atoms is expected on the low-coverage surfaces of the UHV cells, while
the LT cell surfaces with their higher alkali coverage favor the formation of metallic aggregates. 
However, because the surface
conditions of these vapor cells are not well understood, and
comprehensive data of desorption from various types of adsorbates
using a wide range of photon energies is still missing, 
future detailed studies are very desirable. 
Such studies of the surfaces of vapor
cells would contribute substantially to various atomic physics experiments
in which the cell surface plays an important role
in terms of not only LIAD but also, e.g., loss of rare radioactive
atoms on the surface~\cite{Ste94}, surface electrical conductivity~\cite{Bou99}, 
and spin relaxation of gases on the surface~\cite{Ale02,New93}. 

\bibliographystyle{prb}
\bibliography{references}

\begin{references}
\bibitem{comment1}
Strictly speaking, the cell surface is coated with a 
helium film in the case of
the LT cells. 
However, we do not consider that the helium film affects the desorption process as also
claimed in Ref.~\cite{Hat02}.
\end{references}

\end{document}